\documentclass{article}
\usepackage{spconf,amsmath,graphicx}
\usepackage[skip=1pt]{caption}
\usepackage{bm}

\title{Enhancing Patch-Based Methods with
Inter-frame Connectivity for Denoising
Multi-frame Images}
\name{Kireeti Bodduna$^{1,2}$ and Joachim Weickert$^1$}
\address{
\begin{minipage}[b]{0.5\linewidth}
{\centering
$^1$Mathematical Image Analysis Group, \\
Faculty of Mathematics and Computer Science, \\
Saarland University, \\ 66041 Saarbr{\"u}cken, Germany.\\}
\end{minipage}
\begin{minipage}[b]{0.51\linewidth}
\centering
$^2$Electron Microscopy Group, \\ Institute of Biophysics, \\
Johann Wolfgang Goethe University Frankfurt, \\
60438 Frankfurt am Main, Germany. 
\end{minipage}}

\begin{document}
%
\maketitle
\begin{abstract}
The 3D block matching (BM3D) method is among the state-of-art 
methods for denoising images corrupted with additive white 
Gaussian noise. With the help of a novel inter-frame 
connectivity strategy, we propose an extension of the BM3D 
method for the scenario where we have multiple images of the 
same scene. Our proposed extension outperforms all the existing 
trivial and non-trivial extensions of patch-based denoising 
methods for multi-frame images. We can achieve a quality difference 
of as high as 28\% over the next best method without using 
any additional parameters. Our method can also be easily generalised 
to other similar existing patch-based methods.
\end{abstract}
\section{Introduction}
\label{sec:intro}
Denoising images corrupted with Gaussian noise is a classical 
image processing application, on which tremendous amount of 
research has been performed. Generalised K-Means clustering (K-SVD)
\cite{elad2006image}, autoencoders \cite{vincent2010stacked} 
and deep neural networks \cite{xie2012image} are worth mentioning 
works in this field. For the past two decades, 
non-local patch based methods \cite{BCM05,DFKE07,Lebrun2012,LBM2013} 
have been among the state-of-art methods for this application. 
Especially, 3D Block Matching (BM3D) in combination with some 
non-linear transformations and inverse transformations is a widely 
used method for denoising images which are corrupted with a variety 
of noise distributions including Gaussian, Poisson and 
Poisson-Gaussian mixture types of noise \cite{makitalo2011optimal,
makitalo2013optimal, makitalo2014noise, pyatykh2013image}. 
All the above mentioned research work, however, has been performed 
on denoising single images. Denoising multi-frame images is a less 
explored field. Electron microscopy, CT imaging, multi-spectral 
imaging are some of the fields where we encounter the problem of 
obtaining one image from multiple noisy images of the same sample/scene. 
Notable works in the field of multi-frame image denoising include 
non-local patch-based denoising methods \cite{Buades2009,
tico2008multi}, low-rank tensor approximation framework 
with Laplacian scale mixture \cite{dong2015low, dong2018robust}, 
Stein's unbiased risk estimator and linear expansion of threshold 
combining methods \cite{blu2007sure, delpretti2008multiframe}, 
multi-scale sparsity denoising of spectral domain optical 
coherence tomography images \cite{fang2012sparsity}, non-local 
energy functional minimisation approach involving spatio-temporal 
image patches \cite{boulanger2010patch}, low-rank tensor 
modelling based denoising approach \cite{hao2018patch}, 
combining blind source separation and block matching for denoising 
CT images \cite{ hasan2018denoising}, and tensor based modeling
for denoising multi-frame images corrupted with speckle noise 
\cite{zhou2017multi}. 

Three different types of baseline strategies can exist that are 
relevant in extending patch-based methods for denoising multi-frame 
images: First, average the noisy images and then denoise the 
averaged image \cite{Buades2009}. The second is a trivial strategy
introduced by us: Denoise the individual 
images and then average the denoised images. Finally, 
a non-trivial strategy \cite{tico2008multi}: In the initial stage 
of patch-based denoising methods, the group of patches that are 
most similar to a patch are acquired from images other than just 
the reference image also. Once this group of patches are acquired, 
a traditional single image denoising approach is performed. 


\begin{table*}[t]
\setlength{\tabcolsep}{3pt}
\begin{minipage}{0.53\textwidth}
\begin{flushright}
\begin{tabular}{ l r  r r  r}
 \hline 
 Image ($\sigma_\mathrm{noise}$) & BM-1 & BM-2 & BM-3 & BM-M \\
 \hline 

 Bridge (80) & 316.95 & 289.69 & 345.88 & \textbf{266.16}  \\  
 Bridge (100) & 352.16 & 330.92 & 394.19 & \textbf{307.50} \\  
               \vspace{0.5em}
 Bridge (120) & 380.81 & 366.45 & 440.26 & \textbf{345.19} \\   
 Peppers (80) & 83.15 & 66.08 & 91.27 & \textbf{59.20} \\ 
 Peppers (100) & 111.16 & 81.88 & 115.15 & \textbf{74.15} \\ 
  \vspace{0.5em}
 Peppers (120) & 141.97 & 99.85 & 139.50 & \textbf{89.64} \\   
 Lena (80) & 82.18 & 81.11 & 107.10 & \textbf{71.38} \\ 
 Lena (100) & 97.89 & 99.84 & 135.53 & \textbf{87.49} \\  
  \vspace{0.5em}
 Lena (120) & 114.43 & 118.85 & 160.20 & \textbf{107.48} \\   
 House (80) & 66.67 & 74.41 & 96.25 & \textbf{62.60} \\
 House (100) & 83.59 & 93.40 & 123.95 & \textbf{80.53} \\  
 House (120) & 98.37 & 116.31 & 152.82 & \textbf{96.62} \\   
 \hline 
 \end{tabular}
\end{flushright}
\end{minipage}
\hspace{2em}
\vspace{1em}
\begin{minipage}{0.35\textwidth}
 \begin{tabular}{ r  r r  r}
 \hline 
 BM-1 & BM-2 & BM-3 & BM-M \\
 \hline 
  299.72 & 277.90 & 340.38 & \textbf{230.99} \\  
  330.23 & 315.56 & 386.17 & \textbf{268.01} \\ 
  \vspace{0.5em}
  354.58 & 348.86 & 430.78 & \textbf{304.76} \\   
  71.68 & 58.39 & 87.81 & \textbf{47.51} \\ 
  91.63 & 70.70 & 109.38 & \textbf{58.45} \\  
  \vspace{0.5em}
  112.10 & 85.39 & 129.35 & \textbf{69.91} \\   
  72.41 & 72.45 & 102.79 & \textbf{56.41} \\ 
  85.95 & 88.72 & 125.81 & \textbf{70.85} \\ 
  \vspace{0.5em}
  96.82 & 104.29 & 150.64 & \textbf{84.99} \\   
  56.79 & 64.70 & 88.92 & \textbf{48.14} \\ 
  68.97 & 80.96 & 114.43 & \textbf{63.03} \\  
  77.96 & 99.01 & 142.30 & \textbf{75.15} \\   
 \hline
\end{tabular}
\end{minipage}
 \vspace{1em}
 \captionof{table}{MSE values after denoising 5-image (left)
 and 10-image (right) datasets}
 \label{table1}
\end{table*}


{\bf Our Contribution} In this work, our main contribution is a novel 
strategy that differs from all the above mentioned strategies at 
three stages. The first and second differences are at the 3D patch 
grouping stage and the third difference is at the aggregation 
stage of acquiring the final denoised image. These ideas
enhance the denoising quality of the algorithm significantly. 
We perform a detailed and comprehensive evaluation of 
all the above mentioned extensions, specifically for BM3D. 
Such a detailed study of extensions for multi-frame image denoising 
is missing for any of the patch-based methods in the existing works. 

{\bf Paper Structure} In Section \ref{sec:modelling}, we 
introduce the various extensions of BM3D for multi-frame 
images. In Section \ref{sec:expAndDisc}, we present the denoising
results of the experiments performed using various extensions 
of BM3D. We also perform an analysis on why our proposed 
extension outperforms the other existing approaches. 
Finally, in Section \ref{sec:concAndOutlook}, 
we conclude the paper with a summary of our work and 
some ideas for future research.


\begin{figure*}[t]
  \centering
  \includegraphics[width=0.225\linewidth]
  {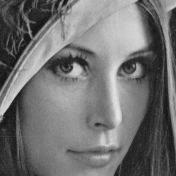}\hspace{0.2em}
  \includegraphics[width=0.225\linewidth]
  {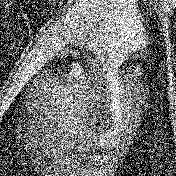}\hspace{0.2em}
  \includegraphics[width=0.225\linewidth]
  {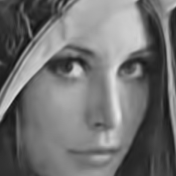}\hspace{0.2em}
  \vspace{0.2em}
  \includegraphics[width=0.225\linewidth]
  {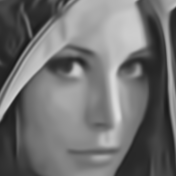}
  \includegraphics[width=0.225\linewidth]
  {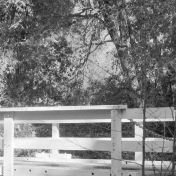}\hspace{0.2em}
  \includegraphics[width=0.225\linewidth]
  {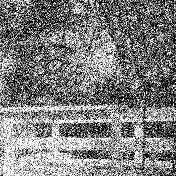}\hspace{0.2em}
    \vspace{0.3em}
  \includegraphics[width=0.225\linewidth]
  {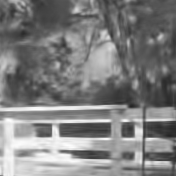}\hspace{0.2em}
  \includegraphics[width=0.225\linewidth]
  					{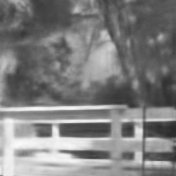}

\caption{Results after denoising 10-image datasets with 
$\sigma_{\mathrm{noise}} = 80$. Top to bottom: Zoomed Lena 
and Bridge images. Left to right: Original, noisy, 
BM3D-M, next best method.}
\label{fig:res}
\end{figure*}


\begin{figure*}[t]
  \centering  					
  \includegraphics[width=0.225\linewidth]
  {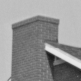}\hspace{0.2em}
  \includegraphics[width=0.225\linewidth]
  {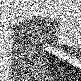}\hspace{0.2em}
  \includegraphics[width=0.225\linewidth]
  {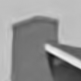}\hspace{0.2em}
    \vspace{0.2em}
  \includegraphics[width=0.225\linewidth]
  {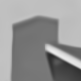}  	
  \includegraphics[width=0.225\linewidth]
  {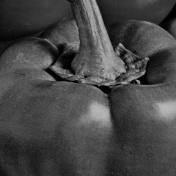}\hspace{0.2em}
  \includegraphics[width=0.225\linewidth]
  {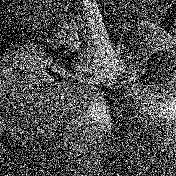}\hspace{0.2em}
  \includegraphics[width=0.225\linewidth]
  {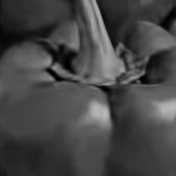}\hspace{0.2em}
  \vspace{0.2em}
  \includegraphics[width=0.225\linewidth]
  {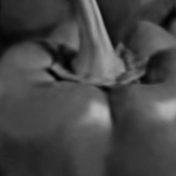}    					
  					
\caption{Results after denoising 10-image datasets with 
$\sigma_{\mathrm{noise}} = 80$. Top to bottom: Zoomed House 
and Peppers images. Left to right: Original, noisy, 
BM3D-M, next best method.}
\label{fig:res1}
\end{figure*}


\section{Modelling of denoising algorithm}
\label{sec:modelling}
\subsection{Original BM3D Method}
BM3D \cite{DFKE07} is a two step method for removing additive 
white Gaussian noise (AWGN). Each step consists of three 
sub-steps. In the first sub-step of the first main step, a 3D 
group is created by finding the most similar patches to a 
reference patch using L$_2$ distance. In the second sub-step, 
a 2D bi-orthogonal spline wavelet transform is applied to this
3D group, followed by a 1D Walsh-Hadamard transform in the third
dimension. After a hard thresholding, the 3D group undergoes 
corresponding 1D and 2D inverse transformations. In the 
third sub-step, a weighted aggregation of all the denoised 
pixels at a particular position in the 2D image domain is 
performed to obtain the final denoised pixel at that particular 
position. Thus, we have a denoised image after the first step. 
In the first sub-step of the second step, the grouping is 
done using the denoised image obtained after the first step.
In the second sub-step, a 2D discrete cosine transform 
followed by a 1D Walsh-Hadamard transform is applied on 
both 3D groups obtained from the noisy image and the denoised 
image from the first step. Then a Wiener filter is applied on 
a combination of both of the above transformed 3D groups. 
The resulting 3D group is then back transformed. In the 
third sub-step, the same corresponding strategy as in the 
first main step gives the final denoised image. 
Thus, the two main steps primarily only differ at the second 
sub-step. We refer the reader to the original work \cite{DFKE07} 
for more details on the method. The final denoised image 
using the weighted aggregation process \cite{Lebrun2012} 
is described as 
\begin{equation}
\label{equation_1}
u^{\textrm{\textbf{final}}}({\bm{x}}) = 
  \frac{\sum\limits_P w_P^{\textrm{\textbf{wien}}} 
  \sum\limits_{Q \in P(\textrm{P})} \chi_Q(\bm{x}) 
  u^{\textrm{\textbf{wien}}}_{Q,P}(\bm{x})}
  {\sum\limits_P w_P^{\textrm{\textbf{wien}}} 
  \sum\limits_{Q \in P(\textrm{P})} \chi_Q(\bm{x}) }.
\end{equation}
Here, $\bm{x}$ is a position in the image domain $Q$ and  
$u^{\textrm{\textbf{final}}}$ is the final denoised image. 
The symbol $u^{\textrm{\textbf{wien}}}_{Q,P}(\bm{x})$  
denotes the estimation of the value at pixel position $\bm{x}$ 
belonging to the patch $Q$ obtained after the Wiener filtering 
of the reference patch $P$. Also, $w_P^{\textrm{\textbf{wien}}}$ 
is obtained from the emperical Wiener coefficients of patch $P$, 
$\mathcal P (P)$ is the set of most similar patches to the 
reference patch $P$, and finally $\chi_Q(\bm{x}) = 1 $ if 
$\bm{x} \in Q$ and 0 otherwise. 


\subsection{Baseline Extensions Applicable for BM3D}
The first extension is a trivial existing one \cite{Buades2009}: 
Average all the noisy frames after aligning them and then 
denoise the averaged image (BM3D-1). Another trivial extension, 
proposed by us, is to denoise every single image and then 
average the aligned denoised images (BM3D-2). A third non-trival 
existing \cite{tico2008multi} extension (BM3D-3) would be to 
first consider a reference image. Then while obtaining a 3D group 
of similar patches in the first step, an L$_2$ distance evaluation 
is performed with patches from all the other images also 
instead of just those from the reference image. 
The rest of the algorithm remains the same. 


\subsection{Our Novel Multi-frame Extension BM3D-M}
Let us now describe the final extension which is 
the main extension proposed by us: 
For every image, we carry out a grouping using L$_2$ distance 
evaluation from patches in all the other images. Thus, we use
the same grouping strategy as the one proposed in BM3D-3, but 
we do this for every image. Then we perform the same filtering 
process as the original BM3D method, for the 3D groups in every 
image as a part of the second sub-step of the first step. In
the third sub-step, the same aggregation process as in the 
original BM3D method is carried out on every image. 
Thus, we have as many denoised images as we have input images. 
In the first sub-step of the second step, we perform the same
grouping strategy as we have in the first sub-step of the 
first step, but using the denoised images of the first step. 
In the second sub-step, we implement the same filtering process 
used in the original BM3D method on 3D groups in every image. 
Finally, in the third sub-step, we carry out an aggregation 
at a particular position in image domain, using a weighted 
aggregation of the denoised pixels obtained in "all the images" 
at this particular position. 


As mentioned in Section \ref{sec:intro}, our method differs 
from the other three strategies in three aspects. First, 
we do not have a reference image like in BM3D-3, but follow the 
denoising procedure for all the images. This is done in order to
maximise the use of available information. 
Second, in the detailed study for selection for BM3D 
parameters \cite{Lebrun2012} and in the original work 
\cite{DFKE07}, two parameters control the number of patches 
in a 3D group: The threshold parameter for L$_2$ distance 
and the maximum number of patches. In this work we do not use the 
former parameter as we have the risk of losing some 
similar patches in the case of highly noisy images. 
Finally, to obtain the final denoised image, all the other 
existing extensions perform the weighted aggregation of pixels 
from 3D groups in just one image. However, 
since we denoise every image unlike the other three strategies, 
we perform the weighted aggregation from denoised 3D groups 
in all the images. The equation that governs the weighted 
aggregation to obtain the final denoised image is modified 
as follows: 
\begin{equation}
u^{\textrm{\textbf{final}}}({\bm{x}}) = 
  \frac{\sum\limits_\ell \sum\limits_{P_\ell} 
  w_{P_\ell}^{\textrm{\textbf{wien}}} 
  \sum\limits_{Q \in P(\textrm{P}_\ell)} \chi_Q(\bm{x}) 
  u^{\textrm{\textbf{wien}}}_{Q,P_\ell}(\bm{x})}
  {\sum\limits_l \sum\limits_{P_\ell} 
  w_{P_\ell}^{\textrm{\textbf{wien}}} 
  \sum\limits_{Q \in P(\textrm{P}_\ell)} \chi_Q(\bm{x}) }.
\end{equation}
In the above equation, we can see the extra summation over all 
the images denoted by index $\ell$. Reference patches $P_\ell$ 
must be considered from every image $\ell$. The rest of the 
symbols have the same meaning as described in \eqref{equation_1}.


\section{Experiments and Discussion}
\label{sec:expAndDisc}
Generally, the various noisy images of the same scene in a 
multi-frame image denoising problem are first registered 
before denoising them to obtain the final image. For the 
registration, one can use a motion estimation method. However, 
in this work, our aim is to carry out a comprehensive 
performance study of the extensions for denoising methods. 
Thus, we assume that all the images are pre-registered 
so that we can straight away compare the denoising qualities 
of various algorithms. It is not difficult to combine the 
denoising strategies proposed by us with state-of-art motion 
estimation methods. Hence, to satisfy this experimental setting, 
we have corrupted Lena, House, Peppers and 
Bridge\footnote{http://sipi.usc.edu/database/} images  with AWGN 
of standard deviations $\sigma_{\mathrm{noise}} = 80, 100, 120$. 
Two datasets for each image for each noise standard deviation 
have been created, one with five realisations of noise and the 
other with ten realisations of noise. 

The threshold parameters for $L_2$ distance while forming the 
3D groups have been excluded in both the main steps of BM3D 
for all the four extensions to keep them on an equal footing.
This particular step is advantageous to all the extensions
for high standard deviation of noise. 
One can introduce back these parameters in applications where there 
is less amount of noise ($\sigma_{\mathrm{noise}}<80$). 
All the other parameters have been chosen according to the detailed 
study of parameter selection for BM3D in \cite{Lebrun2012}. 
It also has to be mentioned that, in the results we showcase shortly, 
we present the mean squared error (MSE) value for BM3D-3 which is 
the best of the 5/10 denoised images obtained when every image is 
selected as the reference frame.

From Table \ref{table1} and Figures \ref{fig:res} and 
\ref{fig:res1}, we can observe that BM3D-M outperforms 
all other methods significantly, in terms of both MSE 
and visually. The images obtained using BM3D-M in particular, 
appear sharper. Moreover, for the Lena 10-image dataset and 
$\sigma_{\mathrm{noise}}=80$, the next best method other than 
BM3D-M is worse by 28\%. Thus, BM3D-M has the capability of 
producing significantly better results than the other algorithms. 
We attribute these results to the crucial differences in 
the modelling: inter-frame connectivity strategy in both main
steps of BM3D to acquire 3D groups and a weighted aggregation 
that connects all the frames instead of just one frame.

It is not clear if the BM3D-1 or the BM3D-2 method is better. 
This is because BM3D-1 does not retain the same explicit noise 
model that was introduced in the original image after averaging
the noisy images and BM3D-2 does not have enough signal in each of the 
input images in each dataset. Also, we would generally expect 
the non-trivial BM3D-3 method to perform better than the BM3D-1 
and BM3D-2 methods. The three algorithms other than BM3D-3 
perform some sort of averaging while BM3D-3 does not. It simply 
does not make full use of the information available in all 
the images. This explains its surprisingly inferior performance.

Experiments on an Intel(R) Core(TM) i7-6700 CPU @3.4 GHz using 
OpenMP indicates that BM3D-M takes 11 times more time 
than single channel BM3D for the 5-image $256\times256$ sized 
House dataset. Also, an ANSI C code combined with a 
CUDA implementation of BM3D-M on an NVIDIA Quadro P5000 architecture 
takes 2.7 seconds. 


\section{Conclusions and Outlook}
\label{sec:concAndOutlook}
We have performed the first comprehensive study of extensions 
for non-local patch-based methods like BM3D for denoising multi-frame 
image datasets. Our proposed extension which uses novel inter-frame 
patch and pixel connectivity strategies, gives significantly better 
results than all the other existing trivial and non-trivial 
extensions. This improvement of the denoising performance is achieved 
without using any additional parameters. The new ideas we introduce can 
also be applied to other similar patch-based methods like BM3D.

In the future, we plan to combine BM3D-M with state-of-art motion 
registration algorithms and also study the error incurred due to false 
registration of pixels. We will also study the performance 
of BM3D-M when the images are corrupted with noise of different standard deviations. Finally, we would apply the algorithm to multi-frame 
bio-physical image datasets obtained from electron microscopes. 

\medskip
{\bf Acknowledgements.}
J.W. has received funding from the European Research Council (ERC)
under the European Union's Horizon 2020 research and innovation programme
(grant agreement no. 741215, ERC Advanced Grant INCOVID).


\bibliographystyle{IEEEbib}
\bibliography{myrefs,refs}
\end{document}